\documentclass[aps,twocolumn,groupedaddress,prb,floatfix,showpacs]{revtex4}
\usepackage{graphicx}
\usepackage{dcolumn}
\usepackage{bm}
\usepackage{amsmath}
\usepackage{verbatim}

\begin{document}

\title{Energy gaps, magnetism, and electric field effects in bilayer graphene nanoribbons}

\author{Bhagawan Sahu$^1$}
\email{brsahu@physics.utexas.edu}
\author{Hongki Min$^2$}
\author{A.H. MacDonald$^2$}
\author{Sanjay K. Banerjee$^1$}
\affiliation{
$^1$Microelectronics Research Center, The University of Texas at Austin, Austin Texas 78758\\
$^2$Department of Physics, The University of Texas at Austin, Austin Texas 78712
}

\date{\today}

\begin{abstract}
Using a \textit{first principles} density functional electronic structure method, we study the energy gaps and magnetism in bilayer graphene nanoribbons as a function of the ribbon width and the strength of an external electric field between the layers.  
We assume AB (Bernal) stacking and consider both armchair and zigzag edges and 
two edge alignments distinguished by different ways of shifting the top layer with respect to the other. 
Armchair ribbons exhibit three classes of bilayer gaps which decrease with increasing ribbon width. 
An external electric field between the layers increases the gap in narrow ribbons and decreases the gap for wide ribbons,
a property which can be understood semi-analytically using a $\pi$-band tight-binding model and perturbation theory. 
The magnetic properties of zigzag edge ribbons are different for the two different edge alignments,
and not robust for all exchange-correlation approximations considered.
Bilayer ribbon gaps are sensitive to the presence or absence of magnetism. 

\end{abstract}

\maketitle

\section{Introduction}

Steady experimental and theoretical progress \cite{allan} in understanding the physics 
of single and multilayer graphene sheets, with and without substrates, \cite{walter} has 
attracted attention from the technical community interested in exploring the 
potential \cite{jeff} of this truly two-dimensional material in electronics. 
Because graphene is atomically thin, it automatically
scales channel-thicknesses to attractive values. 
Isolated single-layer graphene sheets are zero-gap semiconductors when 
extremely weak intrinsic and Rashba spin-orbit interactions (SOIs) \cite{Min}) are ignored,
and have a room-temperature carrier mobility that is weakly \cite{fuhrer,kim} carrier-density dependent 
and higher than in known compound semiconductors.  Because of the relatively weak SOIs, 
and also because the zigzag edges of graphene sheets have been predicted by theory\cite{fujita,louie} to be magnetic, there is also interest in graphene as a potential material for spintronics.
Initial experimental efforts in this direction have focused on injecting spin-polarized carriers from 
magnetic metals. \cite{wees}. 

Before graphene can be used as a replacement for or a supplement to silicon in CMOS circuits, it will be necessary to prepare graphene channels to have a high-resistance or the off state.  The most obvious way to achieve  
off states similar to those of silicon is to induce an energy gap in the density-of-states of a graphene system.
One possibility is to open gaps by tailoring sheet-substrate interactions.  Indeed 
 angle-resolved phtoelectron spectroscopy (ARPES) spectra of single-layer graphene sheets on SiC substrates \cite{lanzara} hint at
this possibility and have been interpreted as showing gaps $\sim $ 0.26 eV.  A more obvious route is to induce sizable quantum-size gaps by using narrow ribbons \cite{nayak}.  
One early experimental report of energy band engineering in graphene nanoribbons \cite{yzhang}
has already appeared in the literature. A second strategy \cite{sahu} for opening gaps in graphene systems is to use a bilayer geometry 
and apply an external electric field directed between the layers. Recent Shubnikov-de Hass cyclotron-mass measurements \cite{Castro} 
and transport studies \cite{oostinga} in bilayer ribbons suggest that such an electric field does indeed open up a gap. The transport measurements suggest that a gap of the order of meV opens up when an electric field of about 0.167 V/nm (which corresponds to applying 50 Volts across a 300 nm SiO$_{2}$ substrate) is applied across the bilayer. A distinct advantage of bilayer ribbons in nanoelectronics was recently suggested by the IBM group \cite{avouris}. Their experiments hint at suppression of electrical noise in bilayer graphene channels 
comapared to the noise present in their single-layer counterparts, thus improving the signal-to-noise ratio. 
Recently, using a tight binding (TB) model \cite{Castro1}, two families of zero-energy edge states were found in bilayer zigzag ribbons.

\begin{figure}[ht]
\includegraphics[width=1\linewidth]{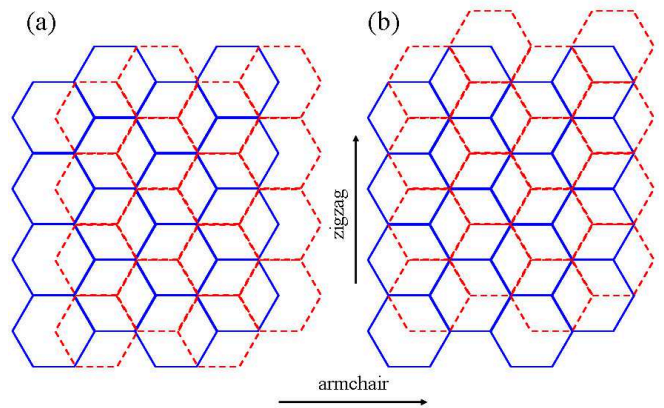}
\caption{ (Color online) Schematic illustration of the two edge alignments we consider in 
bilayer graphene. (a) $\alpha$-alignment and (b) $\beta$-alignment. The $\beta$-alignment can be obtained from the 
$\alpha$-alignment by shifting the top layer.}
\label{fig:Figure1}
\end{figure}

In this article we use density-functional theory \cite{sham} (DFT) based \textit{first principles} methods \cite{gianozzi} and tight-binding (TB) models 
to explore the physics of energy gaps in graphene systems when the narrow ribbon and external field approaches are combined. 
We assume AB (Bernal) stacking and consider two types of edge alignments (Fig. \ref{fig:Figure1}) for both armchair and zigzag ribbons, denoted as $\alpha$ and $\beta$ alignments. (The top layer in $\beta$-alignment is shifted with respect to that in the $\alpha$-alignment.) We report on the ribbon-width and electric field dependence of bilayer gaps for both edge types (armchair and zigzag) and both edge alignments. 
Because zigzag edges in particular have a tendency toward magnetism which, when present, can have a large impact on 
energy gaps, an important element of this study is an assessment of magnetic tendencies in bilayer ribbons.  To the best of our knowledge there are no previous systematic theoretical studies of ribbon-width, magnetism and electric field effects on the energy gaps in bilayer nanoribbons with $\alpha$- and $\beta$-alignments. 

The physics of graphene nanoribbon systems is enriched by the theoretical possibility of 
broken symmetry states.  The ferromagnetism predicted for single layer ribbons with zigzag edges \cite{fujita, louie},
which is related to $\pi$-electron edge orbitals, is one possibility. 
The edge states were seen in non-magnetic STM experiments \cite{kobayashi}. Interestingly half-metallic magnetism has recently been predicted \cite{son} when an electric field is applied across a ribbon with zigzag edges and half-metallicity is argued to be sensitive to the nature of exchange-correlation potential \cite{rudberg}. One \textit{first-principles} DFT study of edge ferromagnetism in bilayer zigzag
ribbons \cite{lee}, which employs the local-density approximation (LDA) \cite{perdew} for the exchange correlation potential,
predicts that a non-magnetic ground state appears when two single-layer graphene systems are stacked.
Our DFT calculations also indicate that edge magnetism is less robust in bilayers than in single-layer systems.
We have found that DFT predictions for the magnetic state of bilayer zigzag ribbon systems, in both edge alignments, are 
sensitive to the particular semi-local approximation that is employed; magnetism appears when the generalized gradient approximations (GGA), 
PW91 \cite{wang} or PBE96 \cite{burke} are used, but not in the LDA. 
Even in single-layer graphene, one recent study\cite{yazyev} has indicated that edge disorder 
strongly influences edge magnetism and hence ribbon gaps.  
The possibility of a novel broken symmetry associated with non-local exchange 
effect, recently predicted using Hartree-Fock theory\cite{min2007},  peculiar to Bernal stacked graphene which would influence ribbon gaps, is not considered 
here since the physics behind it would not be picked up by any general-purpose exchange-correlation approximation.
For all these reasons, the experimental predictions from the present DFT study of bilayer ribbon gaps have considerable uncertainty.  
We believe that the study is nevertheless useful because it can provide a framework for 
assessing the significance of future experimental findings.

Our paper is organized as follows.  We first summarize the DFT calculations that we have 
performed, commenting on the motivation for choosing various different semi-local approximations
for the exchange-correlation potential in section II.  Then, in section III,  we summarize the results that we have
obtained for bilayer ribbon gaps, focusing most extensively on the interplay between
ribbon width, ribbon edge magnetism, and the external electric field between ribbon layers.
Finally we summarize our results and present our conclusions. 

\section{Density Functional Theory Calculations} 

Our electronic structure calculations were performed with plane wave basis sets and ultrasoft pseudopotentials \cite{ultrasoft}. 
As an initial test to reproduce bulk bilayer graphene electronic structure and equilibrium interlayer separation, 
we placed the bulk bilayer graphene in a supercell with 10 {\rm \AA} vacuum regions inserted along the direction perpendicular to the ribbons
to avoid intercell interactions.  The atoms were then fully relaxed without constraints. 
A 21 $\times$ 21 $\times$ 1  {\bf k}-point mesh in the full supercell Brillouin zone (FBZ) was used with a 
30 Rydberg kinetic energy cut-off.  We estimated that these values yield a total energy converged to within 0.01 meV/supercell.
When combined with a LDA, these calculations yielded
an interlayer separation of $\sim 3.34 {\rm \AA}$, within $1\%$ of the experimental interlayer separation (3.35 {\rm \AA}).
The same calculations with PBE96 and PW91 potentials overestimated the equilibrium layer separation rather badly (4.43 {\rm \AA} and 3.65 {\rm \AA} respectively). It is known that LDA or GGA does not include the van der Waals dispersion interactions as a result interlayer binding energy and spacings are not expected to match with experiments \cite{makarov}. The LDA success in predicting the interlayer distance is, therefore, not completely suprising as it has been shown previously \cite{martins} that some fortuitous cancellation errors may be responsible for its success.

For ribbon calculations, we introduced an additional transverse vacuum region of 10 {\rm \AA}. We used 9 $\times$ 3 $\times$ 5 {\bf k}-point mesh in the FBZ and 30 Rydberg kinetic energy cut-offs in all cases. Total energy convergence was tested by using a larger {\bf k}-point mesh, larger vaccum regions, and larger energy cut-off values. The electric field was applied perpendicular to the bilayer ribbons. We chose to apply several values of the electric field, including the value that was used in a recent experiment \cite{oostinga} which was about 0.17 V/nm, and 
up to the maximum value close to the SiO$_{2}$ dielectric breakdown field of 1 V/nm \cite{maria}.
Ribbon widths as large as 5 nm were considered. The $\sigma$-orbitals along the ribbon edges were saturated with atomic hydrogens. 

We do not consider edge roughness in our calculations, although there is a hint of roughness at the atomic scale \cite{berger} 
in epitaxial graphene (grown on SiC substrate) and theoretically, it was shown that considering edge roughness can have considerable effect on the electronic transport in single layer armchair ribbons\cite{basu}.  
We note that alternate edge functionalizations of the bilayers, other than by hydrogen atoms, 
can also change the electronic structure of bilayer zigzag ribbons \cite{katsnelson} because the localized edge states can react with different radicals thus altering the electronic structure of the ribbons.  Our predictions are therefore most relevant to 
nanoribbons cut in a hydrogen environment and without significant edge disorder.

Because we find that the LDA interlayer separation is close to the experimental value, 
we used only the LDA for our armchair ribbon electronic structure calculations. 
We relax the carbon atoms and the C-H distances (initially chosen equal to the C-H bondlength in CH$_4$ molecule) with the force threshold of 0.1 meV/{\rm \AA}. The relaxed interlayer distances in armchair ribbons were found to be very close to the bulk values. 
As in the single-layer case\cite{louie,ordejon}, we identified three classes of armchair ribbons for both types of 
edge alignments.  All bilayer armchair ribbons were found to be non-magnetic, independent of the exchange-correlation approximation.
(We refer to PW91 potential as the GGA below.)  

Zigzag ribbons with $\alpha$-alignment led to non-magnetic ground states (with the initial configurations either as interlayer ferromagnetic or antiferromagnetic) when the LDA was employed whereas a magnetic ground state, which differs from the non-magnetic ground state in total energy 
by only a few tens of meV, was obtained with the GGA. 
For $\beta$-aligned ribbons, both LDA and GGA predict a magnetic ground state although GGA shows a stronger tendency toward magnetism. 
This means within DFT, the occurrence or absence of magnetism in bilayer ribbons is sensitive to the choice of the semilocal approximation.
We stress here the fact that, in both $\alpha$- and $\beta$-aligned ribbons, the total energy difference between interlayer antiferromagnetic and ferromagnetic order is not large enough to call for a distinct magnetic state. Since our focus was to explore the broken symmetry states in zigzag ribbons and there was considerable uncertainty in predicting the magnetic order with GGA, for consistency and comparisons, we chose interlayer ferromagnetic order in both edge alignments and GGA as a semilocal approximation. Since with the relaxation of atoms the bulk interlayer distance was overestimated with GGA compared to the experiment, we chose to fix the interlayer separation and the atomic coordinates at the experimental value for zigzag ribbon calculations. 

We note that the ferromagnetism which occurs in single-layer zigzag ribbons is thought to be related \cite{fujita} to the flat band which appears in TB models and splits when edge magnetic order is allowed. It is not surprising that adding the second layer can destroy the magnetic order since it also splits the flat band. As we show later in this article, a flat band also appear in $\alpha$-aligned ribbons but is shifted from the Fermi level whereas the flat band lies at the Fermi level in $\beta$-aligned ribbons. The position of the flat band with respect to the Fermi level may explain the magnetism in one alignment but not in the other. 

\section{Graphene Bilayer Ribbon Gaps} 

We now present our results for the width, magnetism, and external electric field dependence of the bilayer gaps in ribbons with both edge type and the edge alignment. Wherever possible we will compare the GGA zigzag ribbon results with corresponding LDA results.

\begin{figure}[ht]
\scalebox{0.4}{\includegraphics{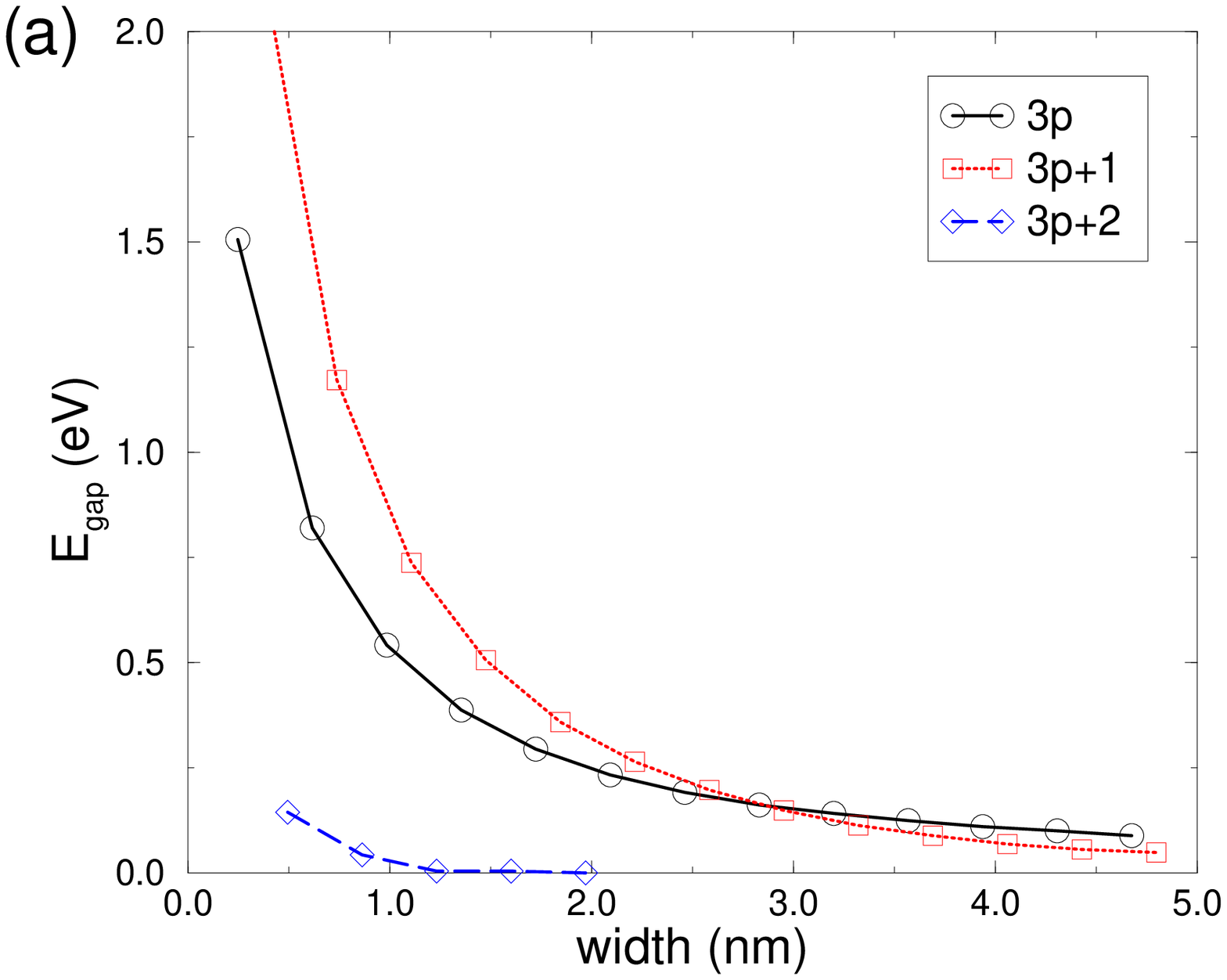}}
\scalebox{0.4}{\includegraphics{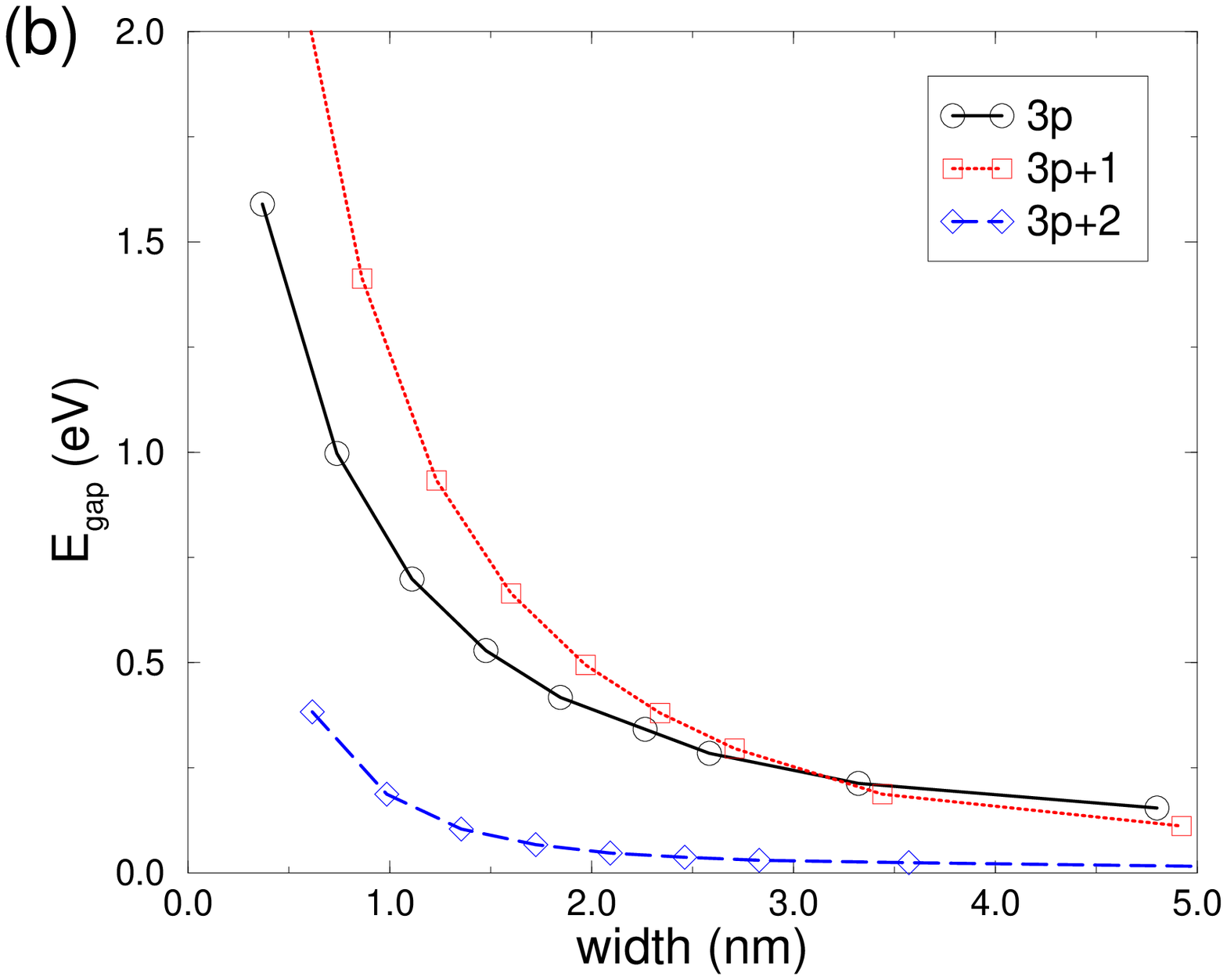}}
\caption{ (Color online) Variation of energy gap with ribbon for balanced bilayer armchair ribbons. 
(a) $\alpha$-alignment and (b) $\beta$-alignment. Three classes of ribbons, denoted as 3{\it p}, 3{\it p}+1 and 3{\it p}+2 are shown. 
For $\alpha$-aligned ribbons, {\it p} = 1-13 whereas for $\beta$-aligned ribbons, {\it p}=1-7, 9, 13}.
\label{fig:Figure2}
\end{figure}
 
\subsection{Balanced Armchair Bilayers}

In this section, we discuss the width dependence of gaps in armchair ribbons with both edge alignments. Figure \ref{fig:Figure2}(a) shows variations of the gap versus the ribbon width for $\alpha$-aligned armchair ribbons.  
Here we label the classes by {\it N} = 3{\it p}, 3{\it p}+1 and 3{\it p}+2 where {\it p} range from 1 to 13 which translates to ribbons with widths of 1-5 nm. As expected on the basis of previous work, the ribbon-width dependence is smooth within three classes which are distinguished by the number $N$ of carbon chains across the ribbon mod 3, with two classes showing semiconducting and class showing a tendency towards metallic behavior. 
The metallic behavior does not appear in the corresponding single layer graphene DFT calculation \cite{louie}. 
We believe that the crossings of 3{\it p} and 3{\it p}+1 curves may be due to the inability of DFT (with LDA or GGA) to predict the gaps accurately especially for narrow gap ribbons. The metallic behavior in 3{\it p}+2 ribbons may be ascribed, similarly, to DFT-LDA not being able to resolve extremely small gaps. As in the monolayer ribbons \cite{louie}, the gaps decrease with increasing width. 

For comparison and for illustrative purpose, we chose {\it p} = 3-7, 9, 13 for the $\beta$-aligned ribbons. DFT again predicts 
three classes of gaps  (Fig. \ref{fig:Figure2}(b)), with no metallic regime for wider 3{\it p}+2 ribbons.
The gaps are found to be consistently larger compared to the gaps in $\alpha$-aligned ribbons. To give a typical example, the gap in a $\beta$-aligned ribbon with the width of 3.32 nm (0.213 eV) is 50 $\%$ larger than the gap in the $\alpha$-aligned ribbon with the width of 3.20 nm (0.141 eV), although both contain N= 27 chains. 
Eventually, for very wide ribbons the bulk limit of a zero-gap semiconductor is approached.
In comparison, the corresponding DFT energy gap for a monolayer ribbon with the width of 3.2 nm is about 0.3 eV \cite{louie}. 
The energy gaps of bilayer nanoribbons are in general smaller than those of monolayer nanoribbons due to the interlayer coupling.

\begin{figure}[ht]
\scalebox{0.35}{\includegraphics{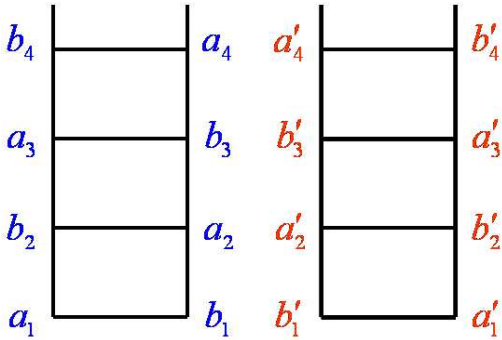}}
\caption{(Color online) Schematic illustration of the TB Hamiltonian
for the coupled two-leg ladder system.  The labels used here are explained in the text.}
\label{fig:Figure3}
\end{figure}

To understand qualitatively the origin of three classes of bilayer armchair nanoribbons in $\alpha$-alignment, we analytically solved for the energy eigenstate  by transforming the Hamiltonian to the one-dimensional TB model of a coupled two-leg ladder system (Fig.\ref{fig:Figure3}).
We calculate the bands only at the ribbon $\Gamma$-point - at which all gaps are minimized.

We made two assumptions: (1) only nearest neighbor (NN) intralayer hopping {\it t} and (2) the NN interlayer hopping $t_\perp$ is allowed.
For $\beta$-alignment this analysis can not be performed due to the dangling bond present with this
particular edge alignment, but we believe that $\beta$-alignment qualitatively follows the same tendency as the $\alpha$-alignment.

In Figure \ref{fig:Figure3}, a solid line represents NN intralayer hopping $t$. 
We consider only NN interlayer hopping $t_{\perp}$,
which links bottom layer $b_i$ sites to top layer $b_i'$ sites for $i=1,2,\cdots,N$.
Then the TB model gives
\begin{eqnarray}
\label{eq:coupled_ladder}
\varepsilon a_n  &=& t( b_{n-1} + b_{n+1} ) + t b_n , \\
\varepsilon a_n' &=& t( b_{n-1}'+ b_{n+1}') + t b_n', \nonumber \\
\varepsilon b_n  &=& t( a_{n-1} + a_{n+1} ) + t a_n + t_{\perp} b_n' , \nonumber \\
\varepsilon b_n' &=& t( a_{n-1}'+ a_{n+1}') + t a_n'+ t_{\perp} b_n  . \nonumber
\end{eqnarray}

Let us define $\alpha_n^{\pm}=(a_n\pm a_n')/\sqrt{2}$
and $\beta_n^{\pm}=(b_n\pm b_n')/\sqrt{2}$.
Then Eq.(\ref{eq:coupled_ladder}) can be rewritten as
\begin{eqnarray}
\label{eq:coupled_ladder_rewrite}
\varepsilon \alpha_n^{\pm} &=& t( \beta_{n-1}^{\pm}  + \beta_{n+1}^{\pm}  ) + t \beta_n^{\pm}  , \\\varepsilon \beta_n^{\pm}  &=& t( \alpha_{n-1}^{\pm} + \alpha_{n+1}^{\pm} ) + t \alpha_n^{\pm} \pm t_{\perp} \beta_n^{\pm}. \nonumber\end{eqnarray}

Assuming $\alpha_n^{\pm}\sim A^{\pm} e^{i n\theta}$ and $\beta_n^{\pm}\sim B^{\pm} e^{i n\theta}$,
we get
\begin{equation}
\left(
\begin{array}{cc}
\varepsilon        & -2 t \cos\theta-t         \\
-2 t \cos\theta-t  & \varepsilon \mp t_{\perp} \\
\end{array}
\right)
\left(
\begin{array}{c}
A^{\pm} \\
B^{\pm} \\
\end{array}
\right)
=0.
\end{equation}
Thus energy spectrum is given by
\begin{equation}
\varepsilon_r^{\pm,\pm}= \pm(t_{\perp}/2) \pm \sqrt{(t_{\perp}/2)^2+(2t\cos\theta_r+t)^2}
\end{equation}
where the edge boundary condition results $\theta_r=r\pi/(N+1)$ for $r=1,2,\cdots,N$.
Note that when $N=3p+2$, there are zero-energy states
$\varepsilon_r^{-,+}=\varepsilon_r^{+,-}=0$ at $r=2(p+1)$.
This means that as in the monolayer case,
bilayer armchair graphene nanoribbons are metallic for $N=3p+2$,
whereas for $N=3p$ and $N=3p+1$, they are semiconducting.

\subsection{Unbalanced Armchair Bilayer Ribbons}

\begin{figure}[ht]
\includegraphics[width=0.8\linewidth]{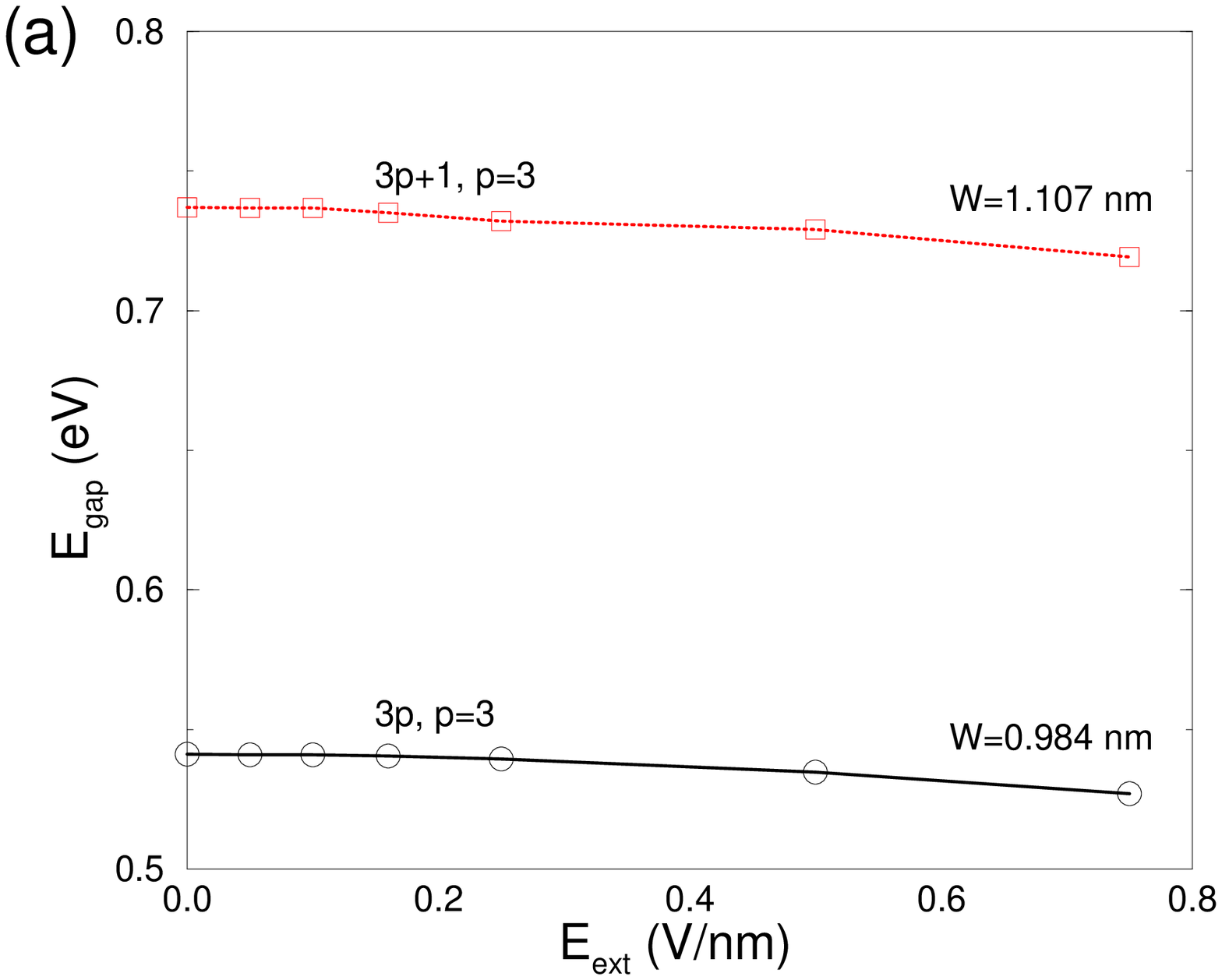}
\includegraphics[width=0.8\linewidth]{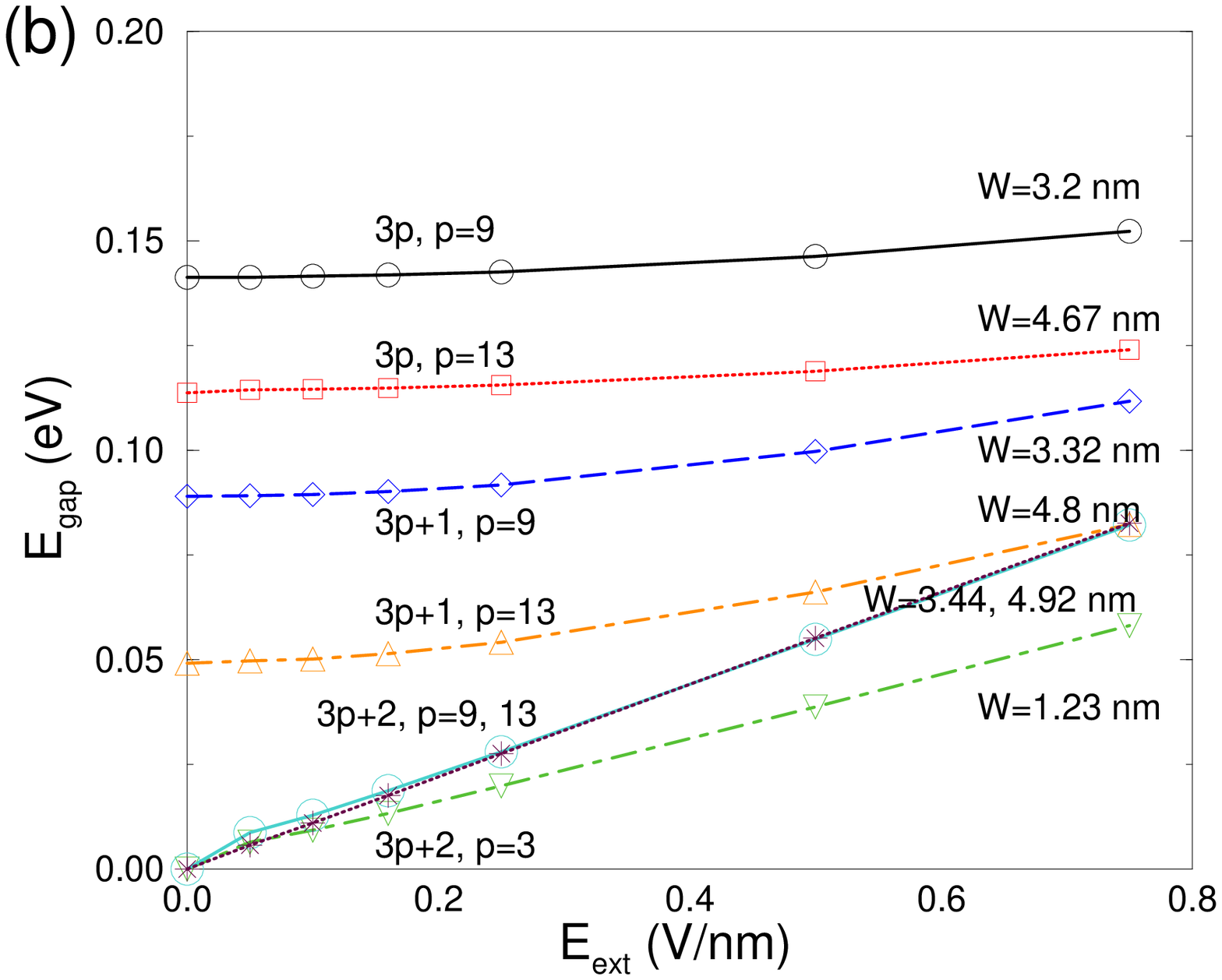}
\caption{ (Color online) Variations of energy gap of the three classes of bilayer armchair ribbons with applied electric field and widths between 1-5 nm for $\alpha$-aligned ribbons.  
For clarity, the ribbon with the width of 4.92 nm is denoted by stars
}
\label{fig:Figure4}
\end{figure}

For the three classes of the armchair ribbons, in both edge alignments, we now discuss the external field effect on the gaps. Interestingly, we find that for $\alpha$-aligned ribbons with gaps below $\sim$0.2 eV, the electric field has the effect of increasing the gap whereas for those above $\sim$0.2 eV, the gap decreases with electric field, as shown in Figure \ref{fig:Figure4}. This can be understood by using second order non-degenerate perturbation theory to semiconducting ribbons and first order degenerate perturbation theory to metallic ribbons. Below we show that there exist a critical gap, given by $\varepsilon_{gap}^{c}=[(\sqrt{5}-1))/2] t_{\perp}$, that can explain the electric field effects. With $t_{\perp}$ = 0.34 eV \cite{sahu}, we get $\varepsilon_{gap}^{c}$ = 0.21 eV. 

To understand the influence of an external electric potential on the gaps, 
we consider the external potential difference between the layers denoted as $U$ as a small perturbation to the on-site energies.
Then Eq.(\ref{eq:coupled_ladder}) and Eq.(\ref{eq:coupled_ladder_rewrite})
are modified as
\begin{eqnarray}
\label{eq:coupled_ladder_potential}
(\varepsilon - U/2) a_n  &=& t( b_{n-1} + b_{n+1} ) + t b_n , \\
(\varepsilon + U/2) a_n' &=& t( b_{n-1}'+ b_{n+1}') + t b_n', \nonumber \\
(\varepsilon - U/2) b_n  &=& t( a_{n-1} + a_{n+1} ) + t a_n + t_{\perp} b_n' , \nonumber \\
(\varepsilon + U/2) b_n' &=& t( a_{n-1}'+ a_{n+1}') + t a_n'+ t_{\perp} b_n  , \nonumber
\end{eqnarray}
 and
\begin{eqnarray}
 \label{eq:coupled_ladder_potential_rewrite}
 \varepsilon \alpha_n^{\pm} - \frac{U}{2}\alpha_n^{\mp} &=& t( \beta_{n-1}^{\pm}  + \beta_{n+1}^{\pm}  ) + t \beta_n^{\pm}  , \\
 \varepsilon \beta_n^{\pm}  - \frac{U}{2}\beta_n^{\mp}  &=& t( \alpha_{n-1}^{\pm} + \alpha_{n+1}^{\pm} ) + t \alpha_n^{\pm} \pm t_{\perp} \beta_n^{\pm}. \nonumber
\end{eqnarray}

Assuming $\alpha_n^{\pm}\sim A^{\pm} e^{i n\theta}$ and $\beta_n^{\pm}\sim B^{\pm} e^{i n\theta}$,
we finally obtain the following Hamiltonian problem:
\begin{equation}
\label{eq:hamiltonian}
H=\left(
\begin{array}{cccc}
0                & 2 t \cos\theta+t  & U/2               & 0                 \\
2 t \cos\theta+t &  t_{\perp}        & 0                 & U/2               \\
U/2              & 0                 & 0                 & 2 t \cos\theta+t  \\
0                & U/2               & 2 t \cos\theta+t  & -t_{\perp}        \\
\end{array}
\right)
\end{equation}
in the $(A^{+},B^{+},A^{-},B^{-})$ basis.
W can achieve a qualitative understanding of the dependence of 
bilayer ribbon gaps on an external potential by treating $U$ as a perturbation.

For simplicity, let's define $\alpha=2 t \cos\theta+t$, $2\beta=t_{\perp}$, $\gamma=\sqrt{\alpha^2+\beta^2}$
and $\delta=U/2$.
The unperturbed energy levels from the lowest value
are given by
$\varepsilon_1^0=-\gamma-\beta$ ,
$\varepsilon_2^0=-\gamma+\beta$ ,
$\varepsilon_3^0= \gamma-\beta$ and
$\varepsilon_4^0= \gamma+\beta$.
The change of energy levels can be obtained by
second-order perturbation theory as follows:
\begin{eqnarray}
\delta \varepsilon_1&=&-\delta \varepsilon_4=-\left( \frac{\alpha^2}{\beta}+ \frac{\beta^2}{\gamma+\beta}\right) \frac{\delta^2}{2\gamma^2}, \\
\delta \varepsilon_2&=&-\delta \varepsilon_3=-\left( \frac{\beta^2}{\gamma-\beta}- \frac{\alpha^2}{\beta}\right) \frac{\delta^2}{2\gamma^2}. \nonumber
\end{eqnarray}
If these energy levels are low-energy states near the Fermi energy,
the unperturbed energy gap is $\varepsilon_{gap}^{0}=2(\gamma-\beta)$
and the energy gap change due to the perturbation is given by
\begin{equation}
\label{eq:gap}
\delta \varepsilon_{gap}=\left( \frac{\beta^2}{\gamma-\beta}- \frac{\alpha^2}{\beta}\right) \frac{\delta^2}{\gamma^2}.
\end{equation}
Alternatively, we can rewrite Eq.(\ref{eq:gap}) using $\varepsilon_{gap}^{0}$, $U$ and $t_{\perp}$ as
following form:
\begin{equation}
\label{eq:gap_rewrite}
\delta \varepsilon_{gap}= C ( \varepsilon_{gap}^{c}-\varepsilon_{gap}^{0} ) U^2,
\end{equation}
where $C$ is a positive constant and
$\varepsilon_{gap}^{c}= [(\sqrt{5}-1)/2] \, t_{\perp}$.
Note that if $\varepsilon_{gap}^0 > \varepsilon_{gap}^{c}$, $U$ decreases the energy gap, while
for $\varepsilon_{gap}^0 < \varepsilon_{gap}^{c}$, $U$ increases the energy gap in second order.
Thus an external potential can increase or decrease the energy gap
depending upon the size of the gap in $\alpha$-aligned bilayer armchair graphene nanoribbons.

For metallic armchair nanoribbons, $\alpha=2 t \cos\theta+t=0$ in Eq.(\ref{eq:hamiltonian})
for low-energy states near the Fermi energy, thus zero-energy states are degenerate.
The Hamiltonian in the zero-energy subspace becomes
\begin{equation}
H=\left(
\begin{array}{cc}
0                & U/2         \\
U/2              & 0           \\
\end{array}
\right),
\end{equation}
thus a small perturbation $U$ opens an energy gap linearly as $\delta\epsilon_{gap}=U$.

In summary, we find that for ribbons with gap below $\varepsilon_{gap}^{c}$ (Fig. \ref{fig:Figure4}(b),
the gap increases with the electric field  whereas for those ribbons with the gap above $\varepsilon_{gap}^{c}$, it decreases with electric field (Fig. \ref{fig:Figure4}(a)). We note that the $\beta$-aligned ribbon gap show similar tendencies under interlayer external electric fields (figure not shown). 

\subsection{Balanced Zigzag Bilayer Ribbons}

In this section we address bilayer zigzag ribbons, for which the physics of gaps cannot be 
separated from the physics of edge magnetism. 
With the LDA $\alpha$-aligned ribbons were found to be non-magnetic, using either 
ferromagnetic and antiferromagnetic arrangements as a starting configuration. 
On the other hand non-zero magnetic energy (tens of meV but slightly more than $K_{B}T$), 
was obtained using GGA.  The character of the interlayer magnetic coupling in these solutions 
(ferro or antiferro) could not be determined since the total energies were too close (within $K_{B}T$). 

In our view this signals a situation in which the competition between non-magnetic and magnetic states is too close for reliable DFT predictions. 
For comparison with the $\beta$-aligned ribbons below, we chose solutions with 
ferromagnetic order between the layers and antiferromagnetic order across the layer 
in discussing 
the width and external electric field effects on bilayer gaps. 

\begin{figure}[ht]
\includegraphics[width=0.49\linewidth]{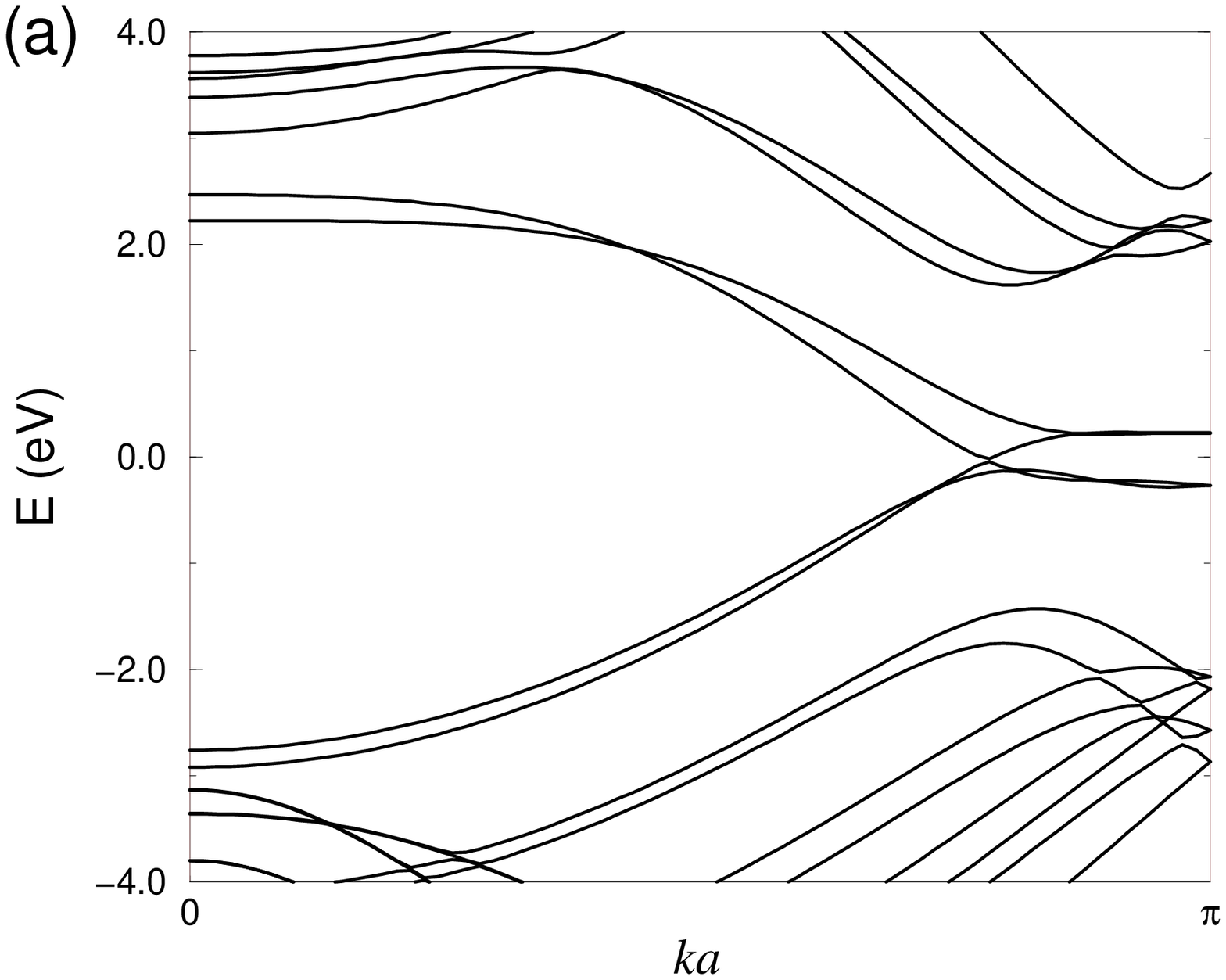}
\includegraphics[width=0.49\linewidth]{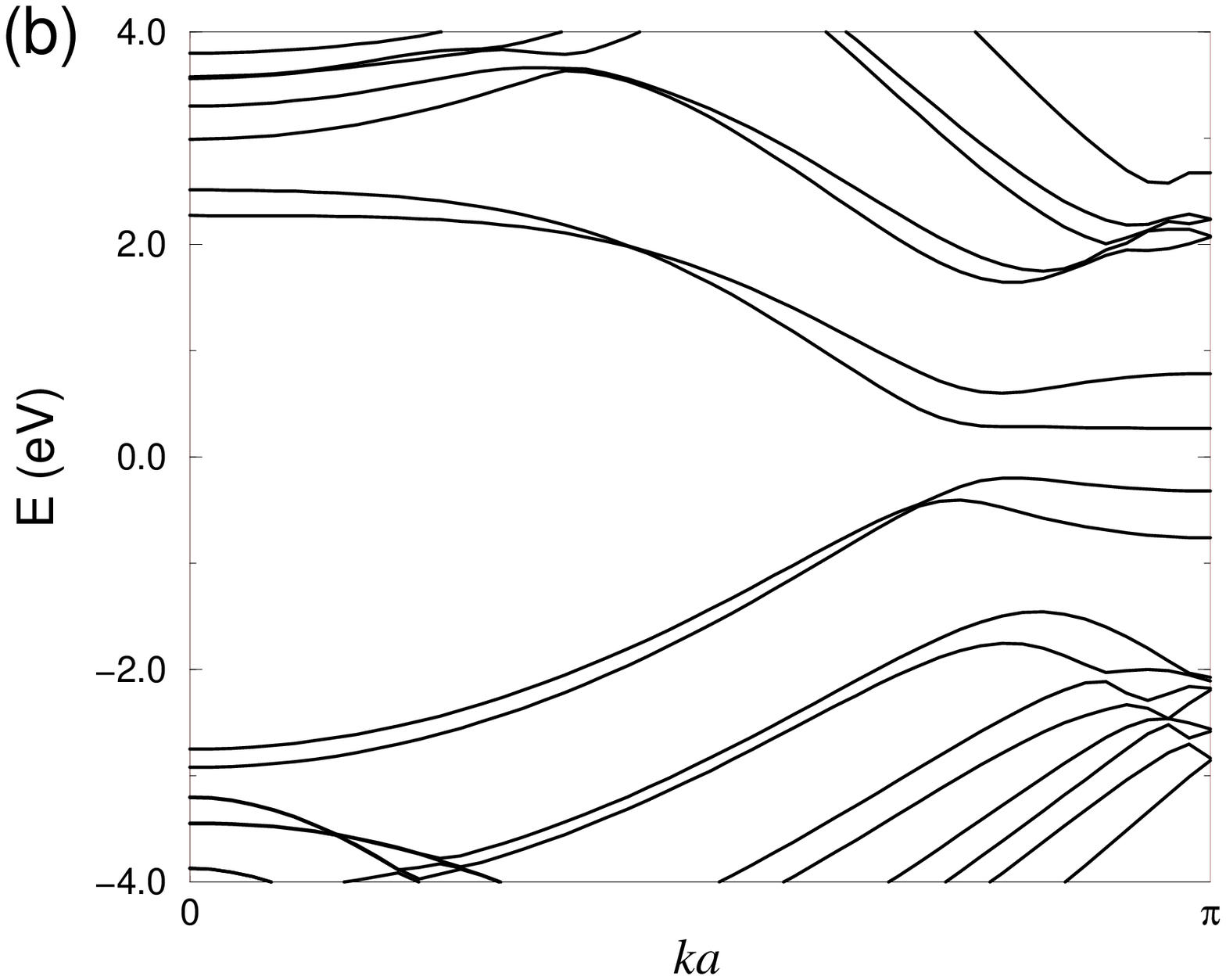}
\caption{Band structure of bilayer zigzag ribbons in $\alpha$-alignment in which the edge atoms are (a) without any magnetic order and (b) with ferromagnetic order between the layers. The width of the ribbon is 1 nm and the Fermi level is set at zero. Clearly the energy spectrum is 
gapless in (a) but has a finite gap in (b).}
\label{fig:Figure5}
\end{figure}

For a typical width of 1 nm, we plot the energy spectrum of the $\alpha$-aligned ribbon (Fig. \ref{fig:Figure5}(a)).  A gapless structure 
with a flat band shifted away from the Fermi level is clearly seen and it occurs roughly at one third the distance from the edge of the BZ. When we allow for magnetic edges in the calculation, due to a different spin order on A and B sublattices along the edges, a gap 
appears at the Fermi level (Fig. \ref{fig:Figure5}(b)). 

\begin{figure}[ht]
\includegraphics[width=0.49\linewidth]{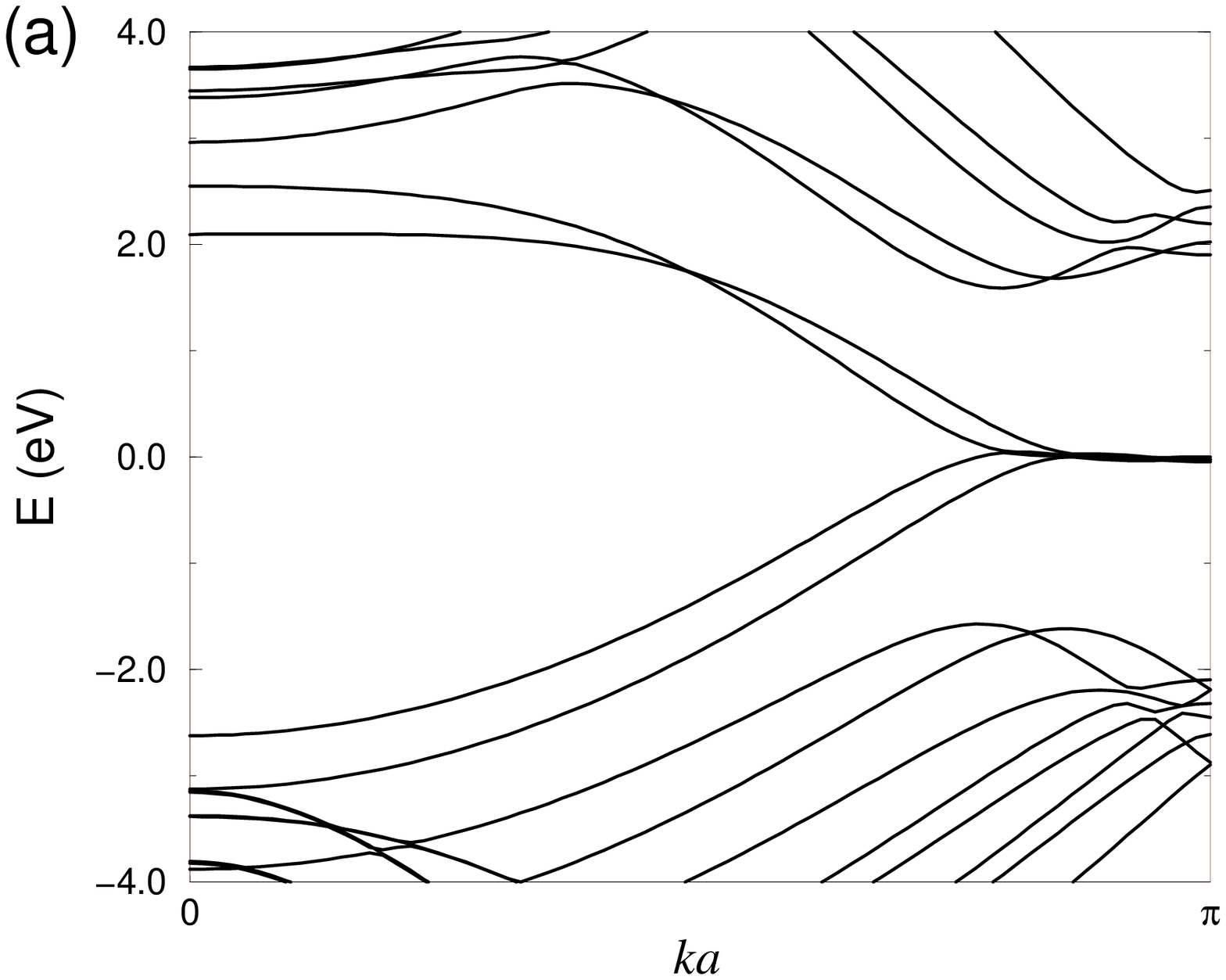}
\includegraphics[width=0.49\linewidth]{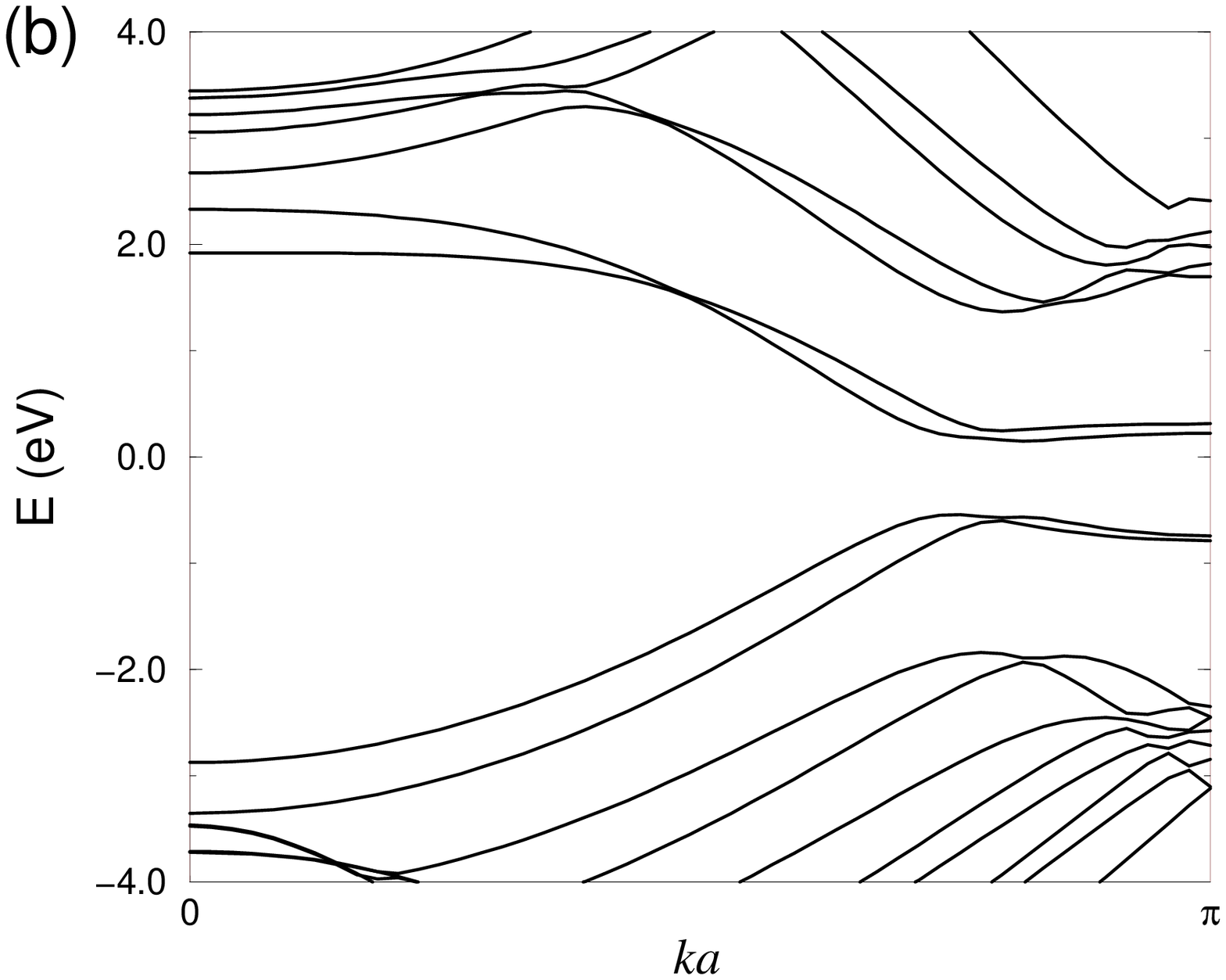}
\caption{ Same as Figure 5 but for $\beta$-aligned ribbons.}
\label{fig:Figure6}
\end{figure}

In $\beta$-aligned ribbons, magnetic ground states were realized with LDA but the magnetic energy was not large enough to distinguish reliably
between non-magnetic and magnetic states. However, GGA predicts a magnetic ground state with a magnetic energy 
of a few hundreds of meV, thus we can say reliably that it is magnetic. Again the precise form of the 
magnetic order was found to be uncertain because of the small energy differences between ferromagnetic and antiferromagnetic 
alignment between the layers. 
The energy band structure for the $\beta$-aligned non-magnetic ribbons is shown in Figure \ref{fig:Figure6}(a). We observe a flat band at the Fermi level which is also seen in single layer graphene calculations. Allowance for magnetic order along the edge atoms lifts the degeneracy 
and a gap opens up in the energy spectrum (Fig. \ref{fig:Figure6}(b)). 

The magnetic versus non-magnetic scenario in $\alpha$- and $\beta$-aligned ribbons respectively can be understood by recognizing the position of the dispersionless state with respect to the Fermi level in the energy spectrum (Figs. \ref{fig:Figure5} and \ref{fig:Figure6}). In $\beta$-aligned ribbons, the occurrence of a flat band at the Fermi level in the non-magnetic ground state results in a large density of states. The system is, therefore, unstable towards developing long-range magnetic order (Stoner's criteria for itinerant magnetism). When exchange interaction is allowed between the edge carbon atoms, the system gains energy by opening up a gap in the energy spectrum and by simultaneously reducing the band energy, thus favoring a magnetic ground state. In $\alpha$-aligned ribbons, in the non-magnetic ground state, the localized state at the Fermi level is absent and therefore the system does not show any clear tendency towards magnetism even when the exchange interactions are included.     

\begin{figure}[ht]
\includegraphics[width=0.8\linewidth]{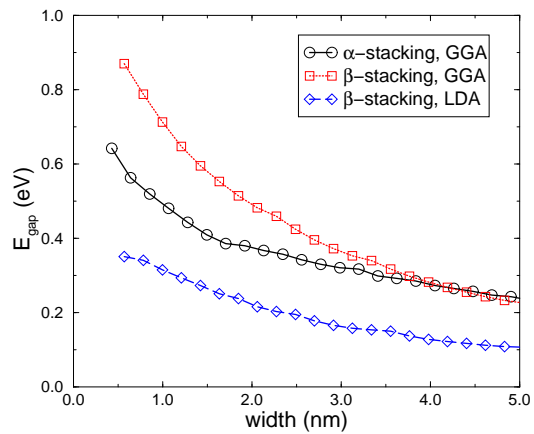}
\caption{ (Color online) Variation of the energy gap with the width of bilayer zigzag ribbons in both edge alignments using both LDA and GGA. 
For $\alpha$-alignments, LDA predicts a non-magnetic ground state and thereofore no energy gap in the energy spectrum.}
\label{fig:Figure7}
\end{figure}

Interestingly, we also found that the magnetic energy is almost independent of ribbon width for both edge alignments. 
This can be understood by analyzing the magnetic moments of the edge atoms for each ribbon width.  We found {\it nearly} the same local moments
($\sim 0.3$ $\mu_{B}$ for $\alpha$-alignment and $\sim 0.32$ $\mu_{B}$ for $\beta$-alignment) on the edge atoms for all widths, and this translates to nearly the same magnetic energy per edge atom (0.2 meV for $\alpha$-alignment and 0.25 meV for $\beta$-alignment) for all ribbon widths.
Since the magnitude of moments on the edge atom is independent of the width of the ribbons, the decrease of the gap with ribbon
width, shown in Figure \ref{fig:Figure7}, can be attributed to weakening quantum confinement effects for the $\pi$-orbitals. 

We note here that for $\beta$-alignment the magnetism is weaker in LDA compared to GGA. 
Also shown in the same figure are the bilayer gaps obtained with the GGA. 
These gaps are consistently larger than those obtained for $\alpha$-aligned ribbons. 
This can be understood by comparing the projected magnetic moments on edge atoms of the $\alpha$- and $\beta$-alignments. 
We found that edge atoms in $\beta$-aligned ribbons carry larger moments than the atoms of $\alpha$-aligned ribbons. 
As a result, the gap is expected to be larger.  In summary, the opening of the gap in the energy spectrum of the magnetic 
ribbons is associated with having opposite magnetic potentials on opposite ribbon edges.
Nevertheless trends in the gap with the width should be understood in terms of quantum confinement effect.  

\subsection{Unbalanced Zigzag Bilayer Ribbons} 

\begin{figure}[ht]
\includegraphics[width=0.8\linewidth]{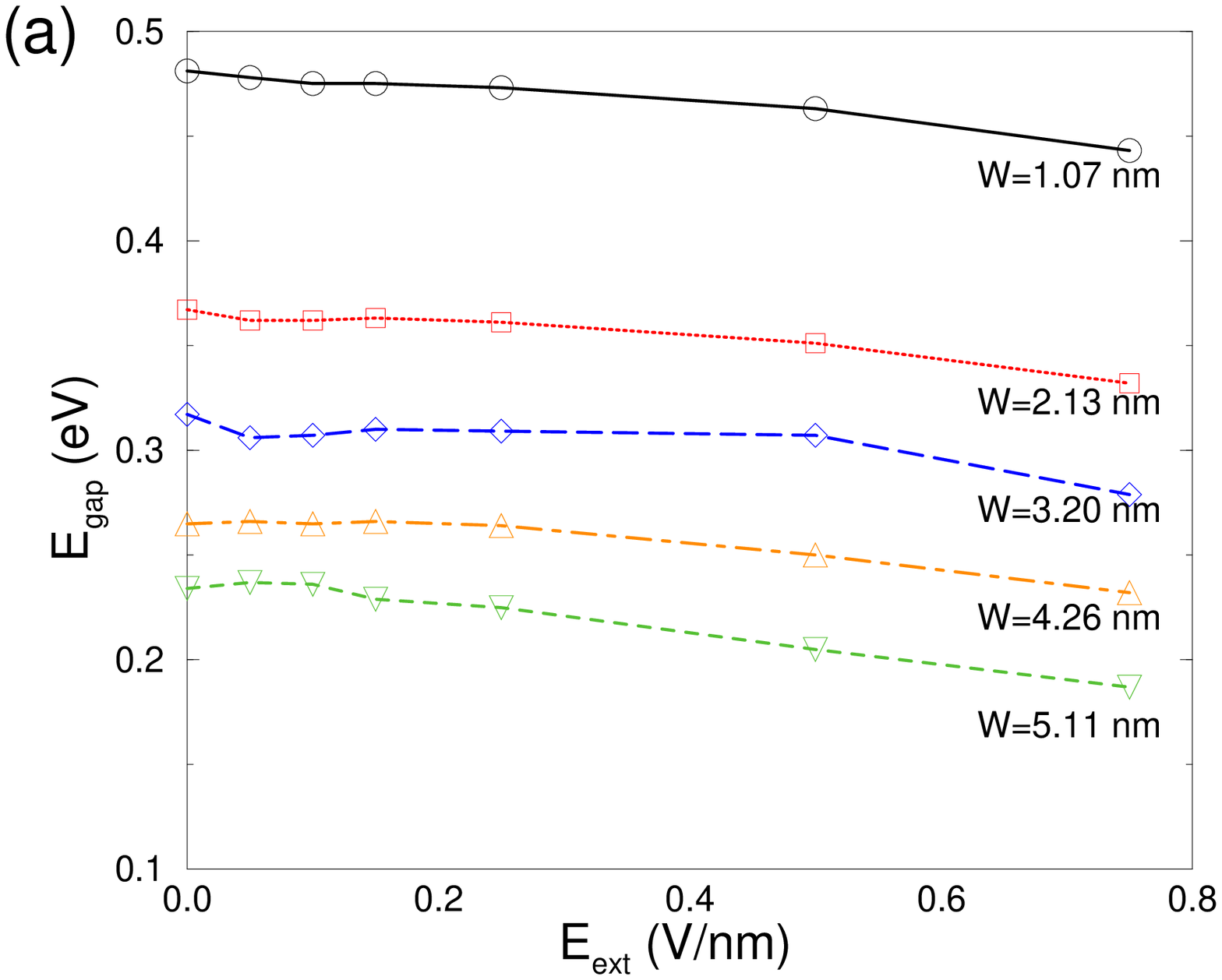}
\includegraphics[width=0.8\linewidth]{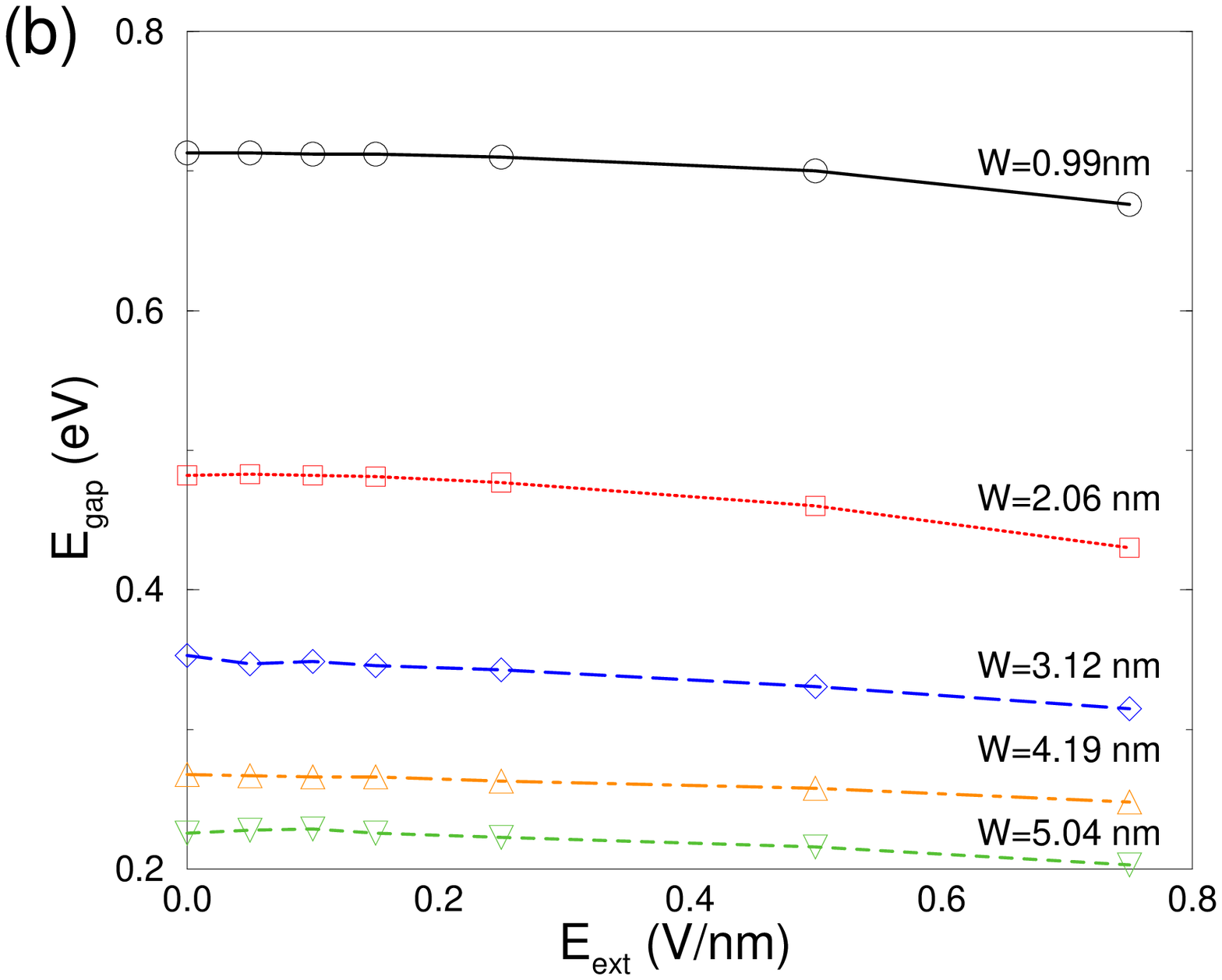}
\caption{(Color online) Variation of the energy gap with external electric field for 
zigzag bilayer ribbons with (a) $\alpha$-alignment and (b) $\beta$-alignment.}
\label{fig:Figure8}
\end{figure}

We apply external electric fields perpendicular to the ribbons with both edge-alignments (up to the dielectric breakdown field of SiO$_{2}$). We find that the gap decreases with increasing electric field in both cases.  Although not explicitly proved by us for zigzag ribbons here, we believe that the gap will decrease with increasing electric fields for zigzag ribbons with widths larger than the critical gap (Fig. \ref{fig:Figure8}(a) and (b)).  

\section{Summary and Conclusions} 

In summary, we have studied the electronic properties of armchair and zigzag bilayer graphene nanoribbons both with and without the external electric fields using a \textit{first principles} DFT-based electronic structure method.  This paper summarizes our results for bilayer ribbons 
with AB (Bernal) stacking with two different edge alignments which we refer to as the $\alpha$- and $\beta$-alignments.
We find three classes of armchair ribbons the origin of which we explained using an analytical TB calculation. 
We discuss the variation of the energy gap with applied electric fields, on the basis of a perturbation theory. 
A critical value of the bulk energy gap exists which controls the sign of the observed behavior.  Gaps
increase (decrease) with electric field for a bulk energy gap below (above) 
the critical value. Magnetic order in zigzag bilayer ribbons is found to be sensitive to the details of the semilocal exchange-correlation
approximation. 
The local moments on edge atoms in zigzag ribbons are found to be very weakly dependent of the width of the ribbon,
which implies that the gap dependence on ribbon width be purely consequence of quantum confinement of $\pi$-orbitals. By invoking band structure effects, we explained the magnetic nature of the these ribbons. 
Although gap values are too sensitive to details of the exchange-correlation potential to allow fully predictive DFT results,
we expect that the present results will 
prove helpful in moving toward a full understanding of how the experimental properties emerge from 
the interplay of magnetic order, width and the external electric field strength.

\acknowledgments

The authors acknowledge financial support from SWAN-NRI and the Welch Foundation. We thank Texas advanced computing center (TACC) for computational support.

\end{document}